\documentclass[aps,pra,superscriptaddress,twocolumn,10pt]{revtex4-1}
\usepackage{amssymb,amsmath,amsfonts,xcolor,graphicx}

\begin{document}

\title{Non-adiabatic holonomic manipulation of the polariton qubit in circuit QED}

\author{Zheng-Yuan Xue}\email{zyxue@scnu.edu.cn} %Corresponding author.
\affiliation{Laboratory of Quantum Engineering and Quantum Materials,  and School of Physics\\ and Telecommunication Engineering, South China Normal University, Guangzhou 510006, China}
%\affiliation{Department of Physics and Center of Theoretical and Computational Physics,  The University of Hong Kong,  Pokfulam Road,   Hong Kong, China}

\author{Wei-Can Yu}
\affiliation{Laboratory of Quantum Engineering and Quantum Materials,  and School of Physics\\ and Telecommunication Engineering, South China Normal University, Guangzhou 510006, China}

\author{Li-Na Yang}
\affiliation{Laboratory of Quantum Engineering and Quantum Materials,  and School of Physics\\ and Telecommunication Engineering, South China Normal University, Guangzhou 510006, China}

\author{Yong Hu}\email{huyong@mail.hust.edu.cn}
\affiliation{School of Physics, Huazhong University of Science and Technology, Wuhan 430074, China}

\date{\today}

\begin{abstract}
As a qubit usually has a limited lifetime, its manipulation should be as fast as possible, and thus  non-adiabatic operation is more preferable. Moreover, as a qubit inevitably  interacts with its surrounding environment, robust operations are of great significance.  Here, we propose a scheme for quantum manipulation of the polariton qubit in circuit QED using non-adiabatic holonomy, which is inherently fast and robust. In particularly, the polariton qubit is shown to be robust against arbitrary low-frequency noise due to its near symmetric spectrum, which can also be convenient manipulated by external microwave driven fields in a holonomic way. Therefore, our scheme presents a promising way of manipulating polariton qubits for on-chip solid-state quantum computation.
\end{abstract}

\pacs{03.67.Lx, 42.50.Dv, 85.25.Cp}

\maketitle

\section{Introduction}
Geometric phases  depend only on certain global geometric property of the evolution path, and thus are largely robust against local noises. This distinct feature makes them promising in implementing quantum computation in a fault-tolerant way \cite{qc} and high-fidelity geometric quantum gates have been experimentally realized \cite{e1,e2}.  Geometric phases can be classified by two types, i.e., the abelian and non-abelian ones, quantum computation based on which are called  geometric \cite{xbw,zhu1,zhu2,xue} and holonomic quantum computation (HQC) \cite{h,h1,h2,duan}, respectively.   The implementation of quantum gates with geometric phases can be achieved by both adiabatic and non-adiabatic evolution. However, the main challenge of the adiabatic method is the long run time, which is comparable with the lifetime of typical qubits \cite{xbw,zhu1}. For this reason, non-adiabatic geometric gates should be more preferable. Recently,  non-adiabatic HQC has been proposed with three-level lambda systems \cite{n1,n2,n3,n4} with several  experimental verifications of elementary gates for universal quantum computation \cite{e3,e4,e5,e6}.  However, in such schemes, the upper excited state is resonantly coupled, and thus the limited lifetime of this state is the main challenge in practical experiments.

Here, we propose to manipulate the polariton qubit in a  non-adiabatic holonomic way. The polariton qubit is choose to be a subset of a V-configuration three-level system,  which is formed by three dressed states in a typical circuit QED setup. Moreover, due to the polaritonic nature of the qubit, the polariton qubit is shown to be robust against arbitrary low-frequency noise. Finally, the non-adiabatic holonomic quantum gate can be constructed by external driven microwave fields. Therefore, our scheme presents a non-adiabatic and geometric way of manipulating polariton qubits for robust on-chip quantum  computation.

\section{The polariton qubit}
We firstly consider implementing the proposed polariton qubit with a typical circuit QED setup \cite{cqed3}, where a superconducting transmon qubit \cite{sq} is capacitively coupled to a one-dimensional transmission line resonator (1D cavity). Setting $\hbar=1$, the interaction Hamiltonian can be written as \cite{cqed2}
\begin{eqnarray}\label{jc}
H_{\text{JC}} = \omega_{\text{a}} |1\rangle\langle1| +\omega_{\text{r}} a^{\dagger}a +g(a\sigma^{+} + a^{\dagger} \sigma^{-}),
\end{eqnarray}
where $\omega_{\text{a}}$ and $\omega_{\text{r}}$ are the frequencies of qubit and cavity, respectively; $\sigma^{-}=|0\rangle\langle1|$ with $|0\rangle$ and $|1\rangle$ being the ground and excited states of the transmon qubit. Label $|n\rangle_{\text{r}}$ as the Fock state of the cavity, the ground state of Hamiltonian (\ref{jc}) is $|G\rangle=|0\rangle|0\rangle_{\text{r}}\equiv|0,0\rangle$ with its eigenvalue set to be $E_{G}=0$. As shown in Fig. 1a, for $n\geq1$, due to the transmon qubit-cavity interaction, each energy level is spitted into two, the eigenstates are
\begin{subequations}
\begin{eqnarray}
{|-,n\rangle}&=\cos\alpha_{n}|0,n\rangle-\sin\alpha_{n}|1,n-1\rangle,\\
{|+,n\rangle}&=\sin\alpha_{n}|0,n\rangle+\cos\alpha_{n}|1,n-1\rangle,
\end{eqnarray}
\end{subequations}
and corresponding eigenvalues are
\begin{eqnarray}\label{E} %\frac{1}{2}
E_{n,\pm} = n\omega_{\text{r}}+{1 \over 2} \left(\delta\pm\sqrt{\delta^2+4ng^2}\right),
\end{eqnarray}
where $\tan(2\alpha_{n})=2g\sqrt{n}/\delta$ with the detuning $\delta=\omega_{\text{r}}- \omega_{\text{a}}$. The transition frequencies are
\begin{subequations}\label{omega}
\begin{eqnarray}
\omega_{n,\pm} &=\omega_{\text{r}}\pm\frac{1}{2}(\sqrt{\delta^2+4(n+1)g^2} -\sqrt{\delta^2+4ng^2}),\\
\omega_{n,\diagup} &=\omega_{\text{r}}+\frac{1}{2}(\sqrt{\delta^2+4(n+1)g^2} +\sqrt{\delta^2+4ng^2}),\\
\omega_{n,\diagdown}&=\omega_{\text{r}}
-\frac{1}{2}(\sqrt{\delta^2+4(n+1)g^2}+\sqrt{\delta^2+4ng^2}),
\end{eqnarray}
\end{subequations}
where $\omega_{n,\pm}$ are the transition frequencies between the  same transmon states, while $\omega_{n,\diagup}$ and $\omega_{n,\diagdown}$ corresponds to the transitions with different transmon states. It is well known that the dynamics of Hamiltonian (\ref{jc}) is constrained within the subspace with certain $|n\rangle_{\text{r}}$. For the zero- and one-excitation subspace,  the three lowest eigenstates $|G\rangle$, $|-,1\rangle$, and $|+,1\rangle$ form a V-configuration, and the latter two are defined as the polariton qubit, as shown in Fig. 1.  For the sake of simplicity, we write the qubit states $|\pm,1\rangle$ as  $|\pm\rangle$ in the following.

\section{Decoherence}
The major decoherence source in superconducting qubits is the low-frequency noise \cite{sq1,sq2,sq3}. The three-level system we considered has better stability under low-frequency noises compared with conventional superconducting qubits. We first consider the influence of the transverse noise
%\begin{eqnarray}
$H^{\prime}= A_x \sigma_1^x$
%\end{eqnarray}
from a superconducting qubit on the polariton qubit.  The low-frequency nature of the noise determines that it cannot resonantly couplings the transitions $|\pm\rangle\leftrightarrow|G\rangle$ of the polariton qubit. Moreover, $\langle\pm|H^{\prime}|\mp\rangle=0$, i.e., it also can not induce qubit states transition. Therefore, the noise can be treated as static fluctuations with perturbation theory and the corrected eigenenergies of the qubit states, up to  second order, are
\begin{eqnarray}\label{B}
\widetilde{E}_{1,\pm}&=E_{1,\pm}+H_{1,\pm}^{\prime}
+\sum_{n}\sum_{a}\frac{{\mid}H_{1,\pm;n,a}^{\prime}\mid^2}{E_{1,\pm}-E_{n,a}}.
\end{eqnarray}
Note that $H_{1,\pm;n,a}^{\prime}=0$ for $n>2$, and thus will greatly simplified the caculation of Eq. (\ref{B}). We rewrite $\widetilde{E}_{1,\pm}=E_{1,\pm}+{\delta}E_{1,\pm}$,
and the correction energy of $E_{1,-}$ and $E_{1,+}$ is calculated to be
\begin{subequations}
\begin{eqnarray}
{\delta}E_{1,-}&=A_x^2(\frac{1}{2\omega_{0,-}}
-\frac{1}{4\omega_{1,-}}-\frac{1}{4\omega_{1,\diagup}})\\
{\delta}E_{1,+}&=A_x^2(\frac{1}{2\omega_{0,\diagup}}
-\frac{1}{4\omega_{1,+}}-\frac{1}{4\omega_{1,\diagdown}}).
\end{eqnarray}
\end{subequations}
They can be represented by $\omega_r$ and $g$ as
\begin{subequations}\label{deltaE}
\begin{eqnarray}
{\delta}E_{1,-}&=\frac{A_x^2}{2}\left[\frac{1}{\omega_r-g}-\frac{1}{\omega_r+g-2g^2/(\omega_r+g)}\right],\\
{\delta}E_{1,+}&=\frac{A_x^2}{2}\left[\frac{1}{\omega_r+g}-\frac{1}{\omega_r-g-2g^2/(\omega_r-g)}\right].
\end{eqnarray}
\end{subequations}
As $g\ll\omega_r$, they can be approximated as
\begin{eqnarray}\label{correction}
{\delta}E_{\pm}^x \approx \mp {A_x^2 g / \omega_{\text{a}}^2},
\end{eqnarray}
and thus the correction of the qubit energy splitting is $\delta E_q^x=2A_x^2g/ \omega_{\text{a}}^2$. Qubit dephasing is determined by this correction and is significantly suppressed by a factor of $\sim A_x^2/ \omega_{\text{a}}^2$, similar to the case of conventional superconducting qubits working at optimal points \cite{op,op2,tian}. However, the optimal points are not immune to longitudinal low-frequency noise, which couples to the qubit through diagonal matrix elements and thus generates a shift of the energy splitting. Fortunately, projecting longitudinal low-frequency noise of the superconducting qubit into our polariton qubit space, it will also be transformed to the transverse type of noise, and thus has minor contribution to the decoherence of our polariton qubit, as the case discussed in the transverse noise. Therefore, the polariton qubit considered here is protected from both  transverse and longitudinal noises. This is in stark contrast to that of conventional superconducting qubits where the first-order perturbation  always plays the leading role. This is ensured by polaritonic nature of the qubit as well as the symmetric  energy spectrum, where although the perturbations may induce the coupling between qubit states and other energy levels, their contribution to the splitting energy are almost cancelled.

\begin{figure}[tbp]\centering
\includegraphics[width=3.4in]{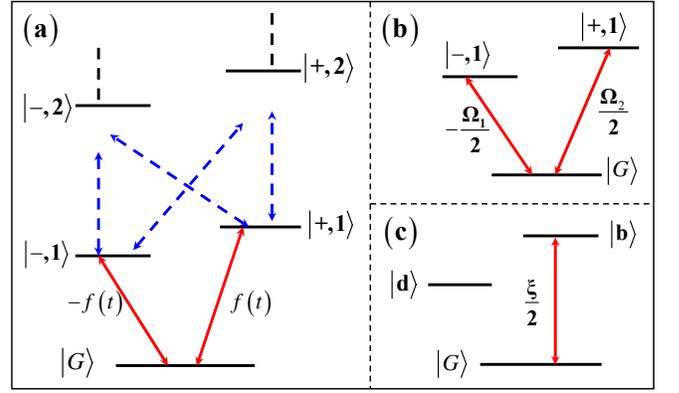} \label{fig1}
\caption{The polariton qubit states. (a) Level structure of the qubit-cavity coupled system. (b) Under approximate parameters of the driven field, the interaction of the lowest V-configuration three levels  can be used to realize single-qubit gates.   (c) The effective interaction in the eigenbasis of the V-configuration levels.}
\end{figure}

\section{Holonomic manipulation}
We now proceed to deal with the manipulation of the polariton qubit with non-adiabatic holonomic gates. Without loss of generality, we set $\delta=0$ in our considered polariton qubit. When a transmon qubit is driven by suitable microwave fields, single-qubit quantum gates for the defined polariton qubit can be induced. Considering that a transmon is driven by a  microwave field as $H_d= \sqrt{2} f(t) \sigma_x$. It is shown that choosing suitable frequencies in $f(t)$, one can control the transitions between between different eigenstates \cite{all}.   Here, we are interested in the three lowest eigenstates, to keep only the couplings between states that form the V-configuration interaction, the driven field  is chosen as
\begin{eqnarray}
f(t) = \Omega_1\cos(\omega_1 t)+\Omega_2\cos(\omega_2 t+\varphi),
\end{eqnarray}
where  $\Omega_n$ is the amplitude of the $n$th component in the driven field with the frequency of $\omega_n$, and $\varphi$ is a prescribed phase difference. Then, we write the full Hamiltonian $H_1=H_{\text{JC}}+H_d$ in the subspace spanned by \{$|G\rangle$, $|-\rangle$, $|+\rangle$\}. Note that this subspace is the eigenspace of $H_{\text{JC}}$, and thus it will be in a diagonalize matrix with the elements being its eigenvalues. Meanwhile, as $\langle\pm|\sigma_x|G\rangle=\langle G|\sigma_x|\pm\rangle=\pm 1/\sqrt{2}$, the matrix elements of $H_d$ will be $\langle\pm|H_d|G\rangle=\langle G|H_d|\pm\rangle=\pm f(t)$. Therefore, the full Hamiltonian can be written as
\begin{eqnarray}\label{h1}
H_1 = \left(
  \begin{array}{ccc}
    E_0 & - f(t) &  f(t) \\
   - f(t) & E_{1,-} & 0 \\
     f(t) & 0 & E_{1,+} \\
  \end{array}
\right).
\end{eqnarray}

In the interaction picture with respect to $H_{\text{JC}}$,   the driven fields can be written as
\begin{eqnarray}
H_{1\text{d}}'=\left(
  \begin{array}{ccc}
    0 & -  f(t) e^{-i\omega_{0,-}t} &  f(t) e^{-i\omega_{0,\diagup}t} \\
        -  f(t) e^{i \omega_{0,-} t} & 0 & 0 \\
    f(t) e^{i\omega_{0,\diagup}t} & 0 & 0 \\
  \end{array} \right). \notag
\end{eqnarray}
Setting $\omega_1=\omega_{0,-}$ and $\omega_2=\omega_{0,\diagup}$, the above  Hamiltonian reduces to
\begin{eqnarray}
H_{1\text{d}}''=\frac{1}{2}\left(
  \begin{array}{ccc}
    0 & -  D_1  &   D_2  \\
        - D_1^* & 0 & 0 \\
    D_2^* & 0 & 0 \\
  \end{array} \right), \notag
\end{eqnarray}
where $D_1\approx \Omega_1 + \Omega_2  e^{i(2gt+\varphi)}\approx  \Omega_1$ and $D_2\approx \Omega_1 e^{-2i gt} +\Omega_2 e^{i\varphi}\approx  \Omega_2 e^{i\varphi}$. The  two step approximations are hold under the justification of the rotating wave approximation. The first one corresponds  the omission of the oscillating terms with much larger frequencies ($\sim 2 \omega_{1,2} \gg 2g$). The second one will be hold when $g \gg (\Omega_1, \Omega_2)$. Therefore, the driven dynamics of the  polariton qubit will be governed by
\begin{eqnarray}
H_{1\text{d}}=\frac{1}{2}\left(
  \begin{array}{ccc}
    0 & - \Omega_1 &  \Omega_2 e^{i\varphi}\\
        - \Omega_1 & 0 & 0 \\
   \Omega_2 e^{-i\varphi}& 0 & 0 \\
  \end{array} \right),
\end{eqnarray}
as shown in Fig. 1b.

In order to investigate the qubit dynamics under the above Hamiltonian, we renormalize it to
\begin{eqnarray}\label{v}
H_{1\text{v}}={\xi \over 2} \left(\sin\frac{\theta}{2}e^{i\varphi} |G\rangle\langle+| - \cos\frac{\theta}{2}|G\rangle\langle-|+ \text{H.c.} \right),
\end{eqnarray}
where the effective Rabi frequency is $\xi=\sqrt{\Omega_1^2+\Omega_2^2}$,  and $\tan(\theta/2)=\Omega_2/\Omega_1$. The eigenstates of the Hamiltonian (\ref{v}) are $|G\rangle$,
\begin{eqnarray}
|b\rangle&=&\sin\frac{\theta}{2}e^{-i\varphi} |+\rangle -\cos\frac{\theta}{2}|-\rangle,\notag\\
|d\rangle&=&\cos\frac{\theta}{2}|+\rangle +\sin\frac{\theta}{2}e^{i\varphi}|-\rangle,
\end{eqnarray}
and thus its dynamics is governed by
\begin{eqnarray}
H_{1}'={\xi \over 2}(|0\rangle\langle b|+ \text{H.c.}),
\end{eqnarray}
which means that, in this eigenbasis, the dynamics of the three level system can be viewed as a resonate coupling between the states $|G\rangle$ and $|b\rangle$ while decouples from the "dark" state $|d\rangle$, as shown in Fig. 1c.

Therefore, the evolution operator  $U =\exp\left(-i\int_0^{t} H_{\text{1v}}dt^\prime \right)$ realizes  the holonomic gate under certain conditions, the reason of which is explained as the following. First, in the qubit subspace, the qubit states evolve according to $|\psi_{\pm}(\tau)\rangle=U|\pm\rangle$.  When the duration of the Hamiltonian should satisfies $\int_0^{\tau}\xi dt=2\pi$, the qubit states  undergo cyclic evolution. Second, at any time, $\langle\psi_a(t)|H_1|\psi_b(t)\rangle=0$ $(a, b\in\{\pm\})$. This justifies that $\{|\psi_+(t)\rangle,|\psi_-(t)\rangle\}$ meets the parallel-transport condition in the qubit states subspace, and thus the evolution is purely geometric. Note that this condition is met due to the structure of the Hamiltonian instead of the slow change of parameters in the adiabatic evolution case. Under the above two conditions, projected the evolution operator onto the qubit space defines the holonomic gate. Therefore, the non-adiabatic holonomic single-qubit gate  can be obtained as
\begin{eqnarray}
U(\theta,\varphi) = \left(
                      \begin{array}{cc}
                        \cos\theta & \sin{\theta}e^{-i\varphi} \\
                        \sin{\theta}e^{i\varphi} & -\cos\theta \\
                      \end{array}
                    \right),
\end{eqnarray}
where $\theta$ and $\varphi$ can be controlled by the driven field, and thus arbitrary single-qubit gate can be obtained by choosing suitable parameters. For example, $\theta=\pi/4$ and $\varphi=0$ implements  the Hadamard gate.

\begin{figure}[tbp]\centering
\includegraphics[width=3in]{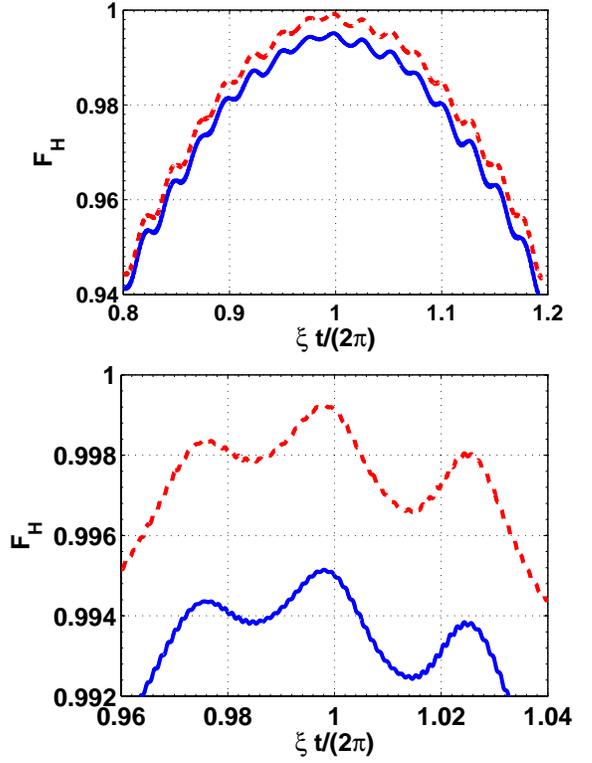} \label{fhgate}
\caption{Fidelity dynamics of the Hadamard gate as a function of $\xi t/(2\pi)$ with the initial state of the polariton qubit being  $|+\rangle$, where the red dashed and blue lines being plotted without and with decohenrence, respectively.}
\end{figure}

The performance of  this gate can be evaluated by considering the influence of dissipation using the  quantum master equation of
\begin{eqnarray}  \label{me}
\dot\rho = i[\rho, H_1]
+\frac \kappa  2  \mathcal{L}(a)  + \frac{\Gamma_1}{2} \mathcal{L}(\sigma^-) + \frac{\Gamma_2}{2}  \mathcal{L}(\sigma_\text{z}),
\end{eqnarray}
where $\rho$  is the density matrix of the considered system, $\mathcal{L}(A)=2A\rho A^\dagger-A^\dagger A \rho -\rho A^\dagger A$ is the Lindblad operator, $\kappa$, $\Gamma_1$  and $\Gamma_2$ are the decay rate of the cavity, the decay and dephasing rates of the qubits, respectively. We consider the Hadamard gate as a typical example and modulate $\xi=g/20$ so that $\xi\ll g$ is well-satisfied to fulfil the requirement of $g \gg(\Omega_1, \Omega_2)$  in the derivation of Hamiltonian in Eq. (\ref{v}). For $\omega_{\text{r}} \simeq 2\pi \times 8$ GHz, the cavity decay rate is $\kappa \simeq 2\pi \times 7$ kHz \cite{cavitydecay} and the capacitive transmon-cavity coupling strength can be $g=\omega_{\text{r}}/20$ \cite{coupling}. As $\theta=\pi/4$ for the  Hadamard gate, $\Omega_1 \simeq 2.41\Omega_2\simeq 0.924 \xi$. Moreover, for a planar transmon qubit couples to a transmission-line resonator can, relaxation and coherence times of 44 and 20 $\mu$s are reported \cite{qubitdecay}, which leads to $\Gamma_1 \simeq 2\pi \times 8$ kHz and $\Gamma_2  \simeq 2\pi \times 3.5$ kHz. As $\kappa$, $\Gamma_1$ and $\Gamma_2$ are all on the same order, for simplicity, we treat them equally as $\Gamma_1=\Gamma_2=\kappa=2\pi \times 8$ kHz. Suppose the qubit is initially in the state of $|+\rangle$,  we evaluate this gate by the fidelity defined by $F_H=\langle\psi_f|\rho|\psi_f\rangle$ with  $|\psi_f\rangle=(|+\rangle+|-\rangle)/\sqrt{2}$ being the ideally final state under Hadamard gate. We have obtained a high fidelity of $F_H \simeq 99.5\%$ at $\xi t/ (2\pi)=1$, as shown in Fig. 2, where the red dashed and blue lines are plotted without and with decohenrence, respectively. Note that to justify our theoretical treatment, we faithfully simulate the quantum process using the original full Hamiltonian $H_1$ in Eq. (\ref{h1}), and thus do not introduce any approximation.

\newpage

\section{Conclusion}

In summary, we propose a scheme for quantum computation with polariton qubit in circuit QED using non-adiabatic holonomy, which is inherently fast and robust. In particularly, the polariton qubit is shown to be robust against arbitrary low-frequency noise due to its near symmetric spectrum, which can also be convenient manipulated by external microwave driven fields in a holonomic way. Therefore, our scheme presents a promising geometric way of manipulating polariton qubits for solid-state quantum computation.

\bigskip
%\noindent\textbf{Acknowledgements}\\

%\acknowledgements
This work was supported by the NFRPC (No. 2013CB921804), the PCSIRT  (No. IRT1243), and the NSFC (Grants No. 11104096, and No. 11374117).

\end{document}